\documentclass[
prl,
twocolumn,
superscriptaddress,
preprintnumbers,
amsmath,
amssymb,
aps,
colorlinks,
urlcolor=blue,
citecolor=blue,
linkcolor=blue
]{revtex4-2}

\usepackage{xcolor}
\usepackage{graphicx}
\usepackage[percent]{overpic}
\usepackage{dcolumn}
\usepackage{bm}
\usepackage[nowatermark]{fixmetodonotes}
\usepackage{orcidlink}
\usepackage{lineno}
\usepackage{isotope}

\begin{document}

\preprint{FERMILAB-PUB-26-0211-CSAID}

\title{Benchmarking State-of-the-Art Theory and Empirical Models of Pionless Neutrino-Argon Scattering in GENIE}

\author{Liang Liu\,\orcidlink{0000-0002-6753-925X}\,}
\affiliation{Fermi National Accelerator Laboratory, Batavia, IL 60510, USA}

\author{Steven Gardiner\,\orcidlink{0000-0002-8368-5898}\,}
\affiliation{Fermi National Accelerator Laboratory, Batavia, IL 60510, USA}

\author{Steven Dytman\,\orcidlink{0000-0002-8278-5299}\,}
\email{dytman@pitt.edu}
\affiliation{University of Pittsburgh, Department of Physics and Astronomy, Pittsburgh, PA 15260, USA}

\date{\today}

\begin{abstract}
Upcoming experiments need improved simulations of neutrino scattering. This work uses the popular GENIE event generator to test a variety of neutrino interaction models against recent MicroBooNE measurements of pionless charged-current interactions. The GENIE code can easily interchange model components, including nucleon form factor parameterizations, quasielastic cross-section calculations, treatments of the nuclear ground state and hadronic final-state interactions. Leveraging this software capability in comparisons with MicroBooNE data, the performance of some of GENIE's most theoretically sophisticated model components is evaluated and contrasted with more empirically-driven alternatives.
\end{abstract}

\maketitle

Neutrino oscillations and cross sections in the GeV region of neutrino energy
are a central part of the frontier of high energy physics~\cite{nustec-review}.
Next-generation experiments, e.g., DUNE~\cite{DUNE:2020lwj,DUNE:2020ypp},
Hyper-Kamiokande~\cite{Hyper-Kamiokande:2018ofw}, and JUNO~\cite{JUNO:2021vlw},
will seek much improved resolution. Present experiments, e.g.,
MicroBooNE~\cite{MicroBooNE:2016pwy}, SBN~\cite{MicroBooNE:2015bmn},
NOvA~\cite{NOvA:2007rmc}, and T2K~\cite{T2K:2011qtm}, are providing data to
move the field forward. Neutrino interaction simulations, e.g.,
GENIE~\cite{genie, Andreopoulos:2009rq, GENIE:2021npt}, must provide accurate descriptions of those
data.

MicroBooNE is currently providing high-quality data for neutrino-argon
scattering that challenge existing cross-section calculations. The mean energy
of the Booster Neutrino Beam used by MicroBooNE is $\sim$~800~MeV. The
experiment is therefore primarily sensitive to nuclear models and reaction
models of quasielastic (QE) and 2 particle--2 hole (2p2h, also commonly labeled
meson exchange current (MEC)) processes.  Full calculations must include pion
production and final state interactions (FSI) of the outgoing hadrons in the
residual nucleus. Because the QE process is difficult to isolate using a
neutrino beam, relevant data are typically called ``QE-like'' and usually
specify a pionless (CC$0\pi$) final state.

GENIE is a leading event generator that provides
interaction simulations for most neutrino experiments. As much as possible,
modern calculations are included, and it has recently obtained important
additions for QE-like interactions. Here, GENIE's latest models are compared
with recent CC$0\pi$ measurements from MicroBooNE~\cite{MicroBooNE:2024yzp,
MicroBooNE:2023tzj,MicroBooNE:2023cmw}. The original MicroBooNE papers already
featured a variety of generator comparisons, and no model was found that could
simultaneously describe all of the data. In this work, more attention is paid
to disentangling the impact of different model ingredients.


Our focus is on lower-energy data where pion production is a minor contribution.  Filali, Munteanu, and Dolan published a similar comparison~\cite{Filali:2024vpy} with MicroBooNE~\cite{MicroBooNE:2023tzj, MicroBooNE:2023cmw} and other data using a combination of the NEUT~\cite{Hayato:2021heg} and NuWro~\cite{Golan:2012rfa} event generators. Focusing on transverse kinematic imbalance (TKI) observables~\cite{Lu:2015hea}, they emphasized the role of nuclear models and had a limited examination of 2p2h and FSI models. Nikolakopoulos \textit{et al.}~\cite{Nikolakopoulos:2024mjj} also focused on TKI variables and studied nuclear models and FSI with the NEUT, NuWro, and ACHILLES~\cite{Isaacson:2022cwh} codes.
Neither of these studies observed satisfactory agreement with MicroBooNE TKI data for various reasons.  Our goal here is to extend these comparisons using GENIE by adding more recent models, doing comparisons within a single program, and broadening the data examined to include a CC$0\pi$ measurement that allows for multiple detected protons~\cite{MicroBooNE:2024yzp}. GENIE has a more flexible structure that features straightforward interchange of model components. This facilitates controlled comparisons in which model ingredients are modified one at a time.
Using both popular empirically-driven GENIE components as well as recently-added models with more sophisticated theory, we examine the sensitivity of the MicroBooNE data to modeling of the nuclear ground state, the nucleon axial-vector form factor, and hadronic FSI.


The AR23 configuration~\footnote{AR23 is an abbreviation of the full GENIE model set
specifier AR23\_20i\_00\_000. This configuration is available in code versions
from v3.4.0 up to the current v3.6.2 release.} of GENIE version 3.6.2 is our starting point. It is presently
used by both SBN and DUNE. For QE-like processes, AR23 model ingredients include the Nieves \textit{et al.} QE~\cite{Nieves:2004wx} and SuSAv2 MEC~\cite{Dolan:2019bxf} cross-section calculations, a $z$-expansion 
axial form factor fit to neutrino-deuterium data~\cite{Meyer:2016oeg} ($F_A^{\rm Deu}$), and a modified local Fermi gas (LFG) nuclear model inspired by spectral function (SF) approaches~\cite{Benhar:1989aw, Benhar:1994hw}. Other relevant processes include pion production, but they contribute less than 14\% of the
MicroBooNE events. This set of models uses the Berger-Sehgal~\cite{Berger-Sehgal}
model for resonance processes and a scaled result from Bodek-Yang~\cite{Bodek:2005de} for nonresonant processes.  FSI are modeled with the hA2018~\cite{Dytman:2011zz} model. The GENIE treatment of pion production
is kept the same throughout this work.

To obtain a good description of the MicroBooNE data, a variety of nuclear
effects must be considered. QE processes depend on a detailed description of
the neutrino-nucleon vertex in the nuclear medium~\cite{nustec-review}. The
full description of that vertex comes from electron scattering (vector
current), and neutrino scattering (adding the axial current). The importance of the nuclear model is
well-established and has been examined many times. Relevant dynamics are
complicated and a variety of solutions have been proposed. Here, we study 
two QE/nuclear models. In GENIE, the Nieves \textit{et al.}
model was adapted to use an LFG
model for scattering from an off-shell nucleon~\cite{GENIE:2021npt}. Although AR23 adds SF-like modifications to the original LFG model~\cite{Nieves:2004wx}, we omit these in all of our other LFG-based GENIE configurations. The SF~\cite{Benhar:1989aw,Benhar:1994hw,Betancourt:2023uxz} model comes from a
description of $(e,e^\prime\,p)$ data and has more sophisticated nuclear
structure. While the SF model provides a joint distribution of nucleon binding
energy and momentum, the Nieves \textit{et al.} model with LFG uses an empirical nucleon momentum and an empirical constant for the binding energy~\cite{Moniz:1971mt}. In this work, a fit to JLab $(e,e'p)$  data~\cite{JeffersonLabHallA:2022cit,JeffersonLabHallA:2022ljj}
is used for the SF nuclear model. We consider two treatments of the nucleon
axial-vector form factor: $F_A^{\rm Deu}$ from AR23 and a result from recent lattice
quantum chromodynamics ($F_A^{\rm LQCD}$) calculations~\cite{Meyer:2026kdl}.
An accurate description of the data also requires a model of
MEC processes. The SuSAv2 model~\cite{Megias:2018ujz} is used in all new comparisons.  Historically, MEC contributions have been hard to separate from axial form factor effects in neutrino measurements.

An FSI treatment is also required to describe the MicroBooNE data. In recent
years, the default GENIE FSI model has been hA2018~\cite{Dytman:2011zz}. It is
a data-based, empirical model that provides a reasonable fit to a variety of
hadron-nucleus data. A recent addition to GENIE is an interface with the more
theoretical Li\`ege intranuclear-cascade model (INCL v6.33.1)~\cite{cugnon:2016ghr,
mancusi:2014eia, Ershova:2023dbv}. This model was developed to describe
hadron-nucleus scattering, especially low-energy nucleon-nucleus processes of
interest for MicroBooNE.



In addition to variations of AR23 in which we change the model components discussed above, we also consider the unaltered AR23
configuration and the MicroBooNE Tune. The MicroBooNE Tune~\cite{MicroBooNE:2021ccs} starts with the G18\_10a\_02\_11a model set from GENIE v3.0.6. This configuration is largely similar to AR23, but it uses the Valencia MEC model, a dipole axial form factor, and an LFG nuclear model without SF-like modifications. Magnitude and shape parameters for QE and MEC interactions were then fit to T2K data~\cite{T2K:2016jor}. The primary fit result is an increase in magnitude of about $12\%$ (QE) and $66\%$ (MEC).

In contrast to AR23 and the MicroBooNE Tune, the remaining configurations of
GENIE considered in this work rely on features not yet in an official release of the code. The SF nuclear model and LQCD axial form factor are anticipated in the upcoming GENIE v3.8.0 with the updated INCL interface in a future release. Our custom
GENIE configurations adopt a version of the Nieves \textit{et al.} QE model that
corrects an implementation error in GENIE v3.6.2 and prior versions.  (See the Supplemental Material~\cite{suppl} for details.) 
This error produces roughly an 8\% increase for the SF QE argon cross section when the Random Phase Approximation (RPA) corrections 
are off (for theoretical consistency) but much smaller for the LFG simulation. RPA is a moderate-sized but poorly understood effect for the MicroBooNE data.  
Because the RPA effects in the MicroBooNE data are  small (less than 10\%)~\cite{suppl}, and 
with the goal of making the most meaningful
comparisons with the SF model, we consistently omit RPA corrections for all
models studied except AR23 and the MicroBooNE Tune. 

In this study, the predictions of GENIE simulations are reported without
variations of model parameters and with the same MEC and pion production models throughout. Apart from AR23 and the MicroBooNE Tune, we
consider all combinations of 2 QE/nuclear model sets, 2 axial form factors, and
2 FSI models (see above for details). Each pair includes one widely-used model
and an alternative that is regarded as more theoretically sophisticated.
Consistency of physics is always a concern. 
With RPA turned off, the Nieves \textit{et al.} and SF QE models~\cite{suppl} are very similar and can be coupled to different nuclear models. Nevertheless, FSI and primary interactions often use different nuclear models.

\begin{figure*}[htbp]

\centering
\begin{minipage}{0.3\textwidth}
\begin{overpic}[width=\linewidth]{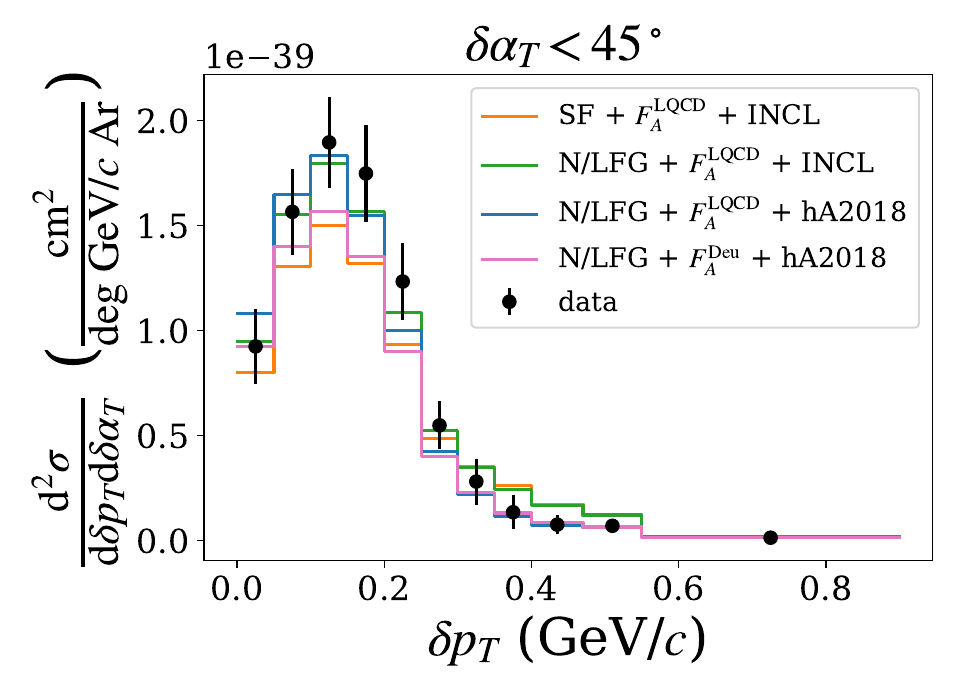}
\put(22,58){\textbf{(a)}}
\end{overpic}
\end{minipage}
\begin{minipage}{0.3\textwidth}
\begin{overpic}[width=\linewidth]{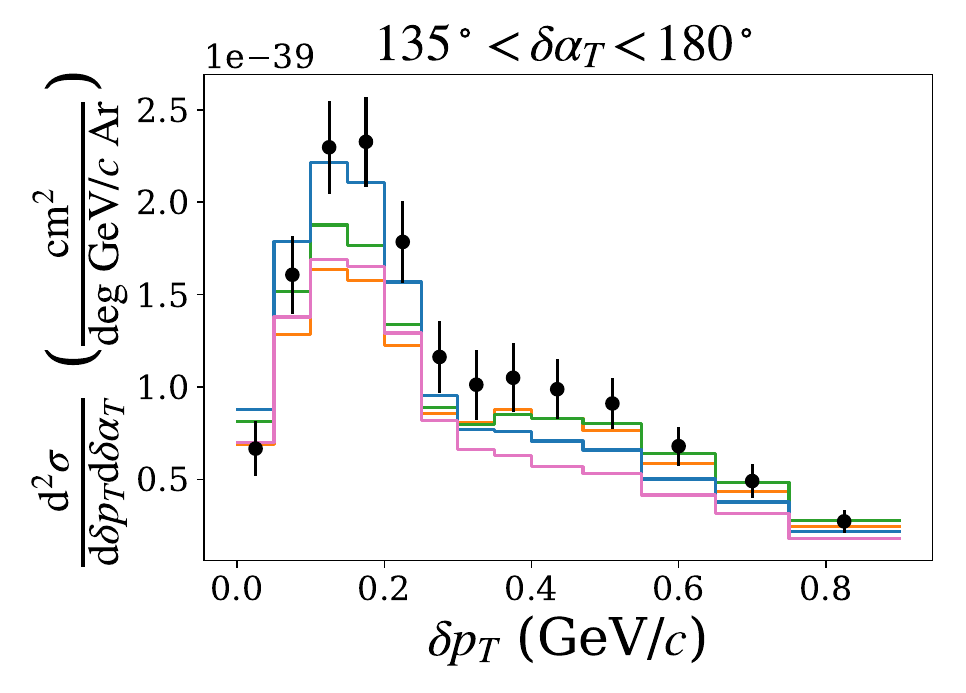}
\put(22,58){\textbf{(b)}}
\end{overpic}
\end{minipage}
\begin{minipage}{0.3\textwidth}
\begin{overpic}[width=\linewidth]{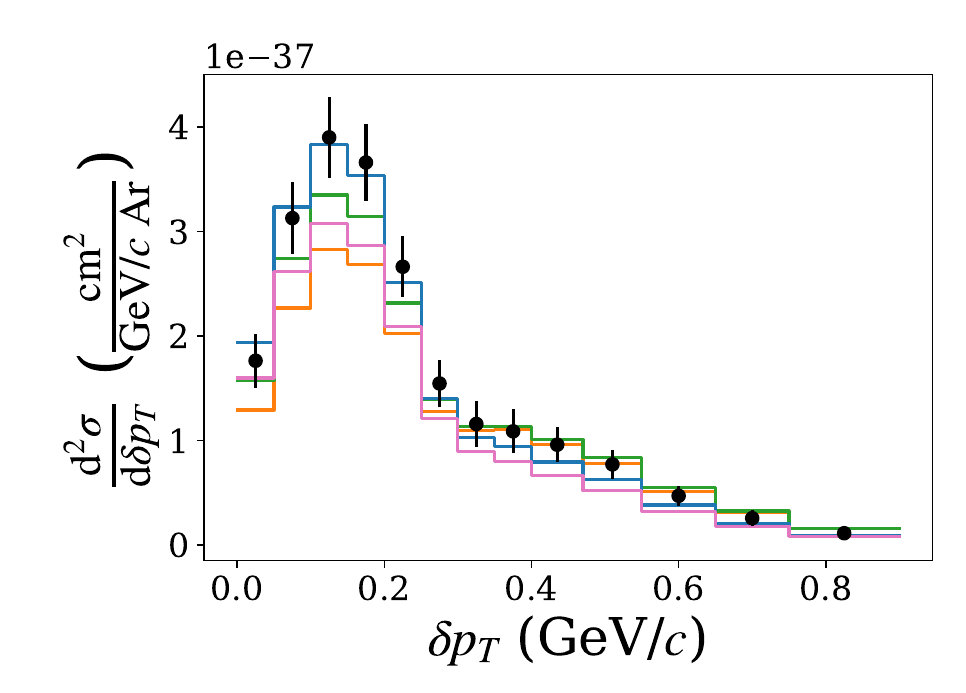}
\put(22,58){\textbf{(c)}}
\end{overpic}
\end{minipage}

\vspace{0.5cm}
\begin{minipage}{0.3\textwidth}
\begin{overpic}[width=\linewidth]{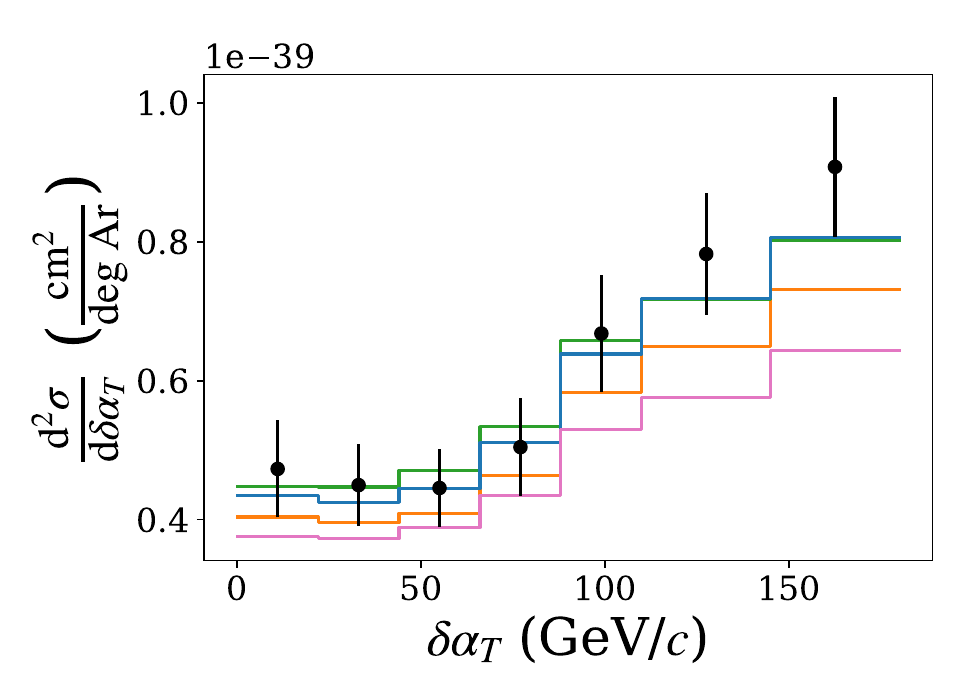}
\put(22,58){\textbf{(d)}}
\end{overpic}
\end{minipage}
\begin{minipage}{0.3\textwidth}
\begin{overpic}[width=\linewidth]{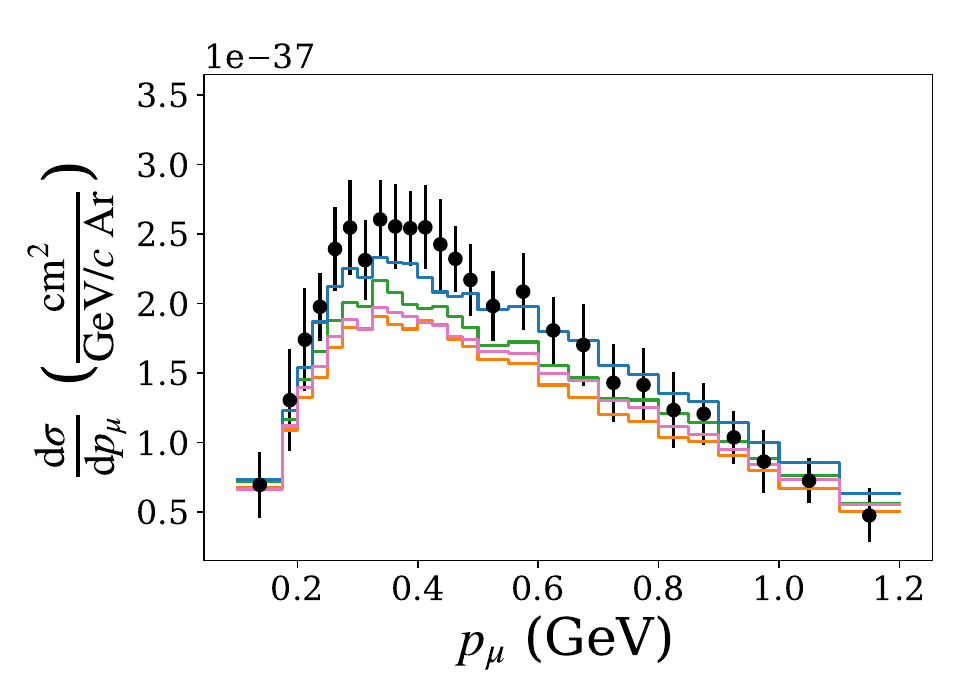}
\put(22,58){\textbf{(e)}}
\end{overpic}
\end{minipage}
\begin{minipage}{0.3\textwidth}
\begin{overpic}[width=\linewidth]{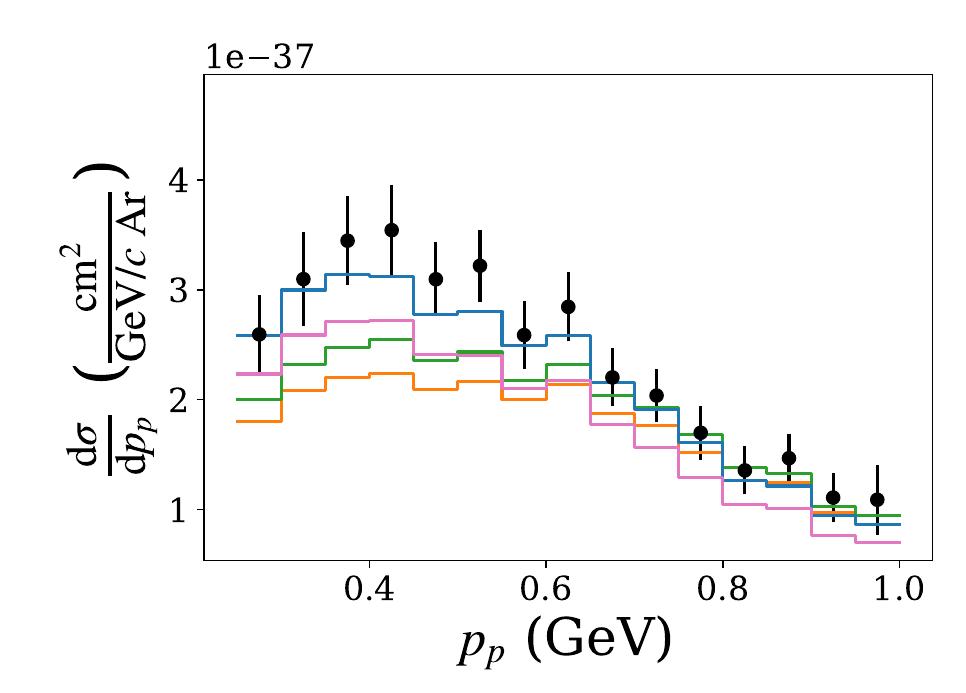}
\put(22,58){\textbf{(f)}}
\end{overpic}
\end{minipage}

\caption{Black points: MicroBooNE data~\cite{MicroBooNE:2023tzj, MicroBooNE:2023cmw,MicroBooNE:2024yzp}.
Panels (a) and (b) show the lowest and highest bins, respectively, of $\delta\alpha_T$ in a two-dimensional measurement with $\delta p_T$ from Ref.~\cite{MicroBooNE:2023cmw}. Panel (c) shows the corresponding single-differential measurement of $\delta p_T$ while panel (d) shows $\delta\alpha_T$. Finally, panels (e) and (f) respectively show the differential cross sections in muon momentum and leading proton momentum from Ref.~\cite{MicroBooNE:2024yzp}.
The primary comparison is with the theory-based model (orange) and various alternate models.  See main text for details.}
\label{fig:theory}
\end{figure*}

Figure~\ref{fig:theory} shows various comparisons with MicroBooNE data.  GENIE's
predictions with the more theoretical models  (the SF
QE/nuclear model, the $F_A^{\rm LQCD}$ axial form factor, and the INCL FSI
model) are shown with the orange histogram. 
The remaining lines are predictions as each
component is progressively swapped with its more empirical counterpart.
Switching to the Nieves \textit{et al.} QE/nuclear model (green) is followed by replacing INCL
with hA2018 for FSI (blue). Finally, the axial form factor taken from
neutrino-deuterium scattering data ($F_A^{\rm Deu}$) is substituted for the
LQCD prediction ($F_A^{\rm LQCD}$), yielding a configuration with all three
empirically-driven components (red). Problems with the most theoretical model
come from both shape and normalization considerations. In particular, its
magnitude is significantly less than data values in all comparisons. In
general, the switches to Nieves/LFG and hA2018 increase the magnitude and the
switch to $F_A^{\rm Deu}$ decreases the magnitude.

\begin{table*}[htbp]

    \caption{$\chi^2/{\rm ndf}$ ($p$-value)  of MC and reference simulations relative to MicroBooNE data. }
    \label{tab:chisq}
    \centering
\begin{ruledtabular}
    \begin{tabular}{lrrrrr}
    Configuration & $\delta p_T$ vs. $\delta \alpha_T$\footnote{The $\chi^2/{\rm ndf}$ ($p$-value) in this column include contributions from intermediate $\delta\alpha_T$ bins not shown in Fig.~\ref{fig:theory}.} & $\delta p_T$ & $\delta \alpha_T$ & $p_{\mu}$ & $p_p$ \\
    \hline
N/LFG + $F_A^{\rm LQCD}$ + hA2018\footnote{Best empirical tune.} & 35.98/49  (0.92) & 6.78/13  (0.91) & 3.47/7  (0.84) & 16.29/26  (0.93) & 16.05/15  (0.38)\\
SF + $F_A^{\rm LQCD}$ + hA2018 & 34.25/49  (0.95) & 6.36/13  (0.93) & 5.19/7  (0.64) & 18.82/26  (0.84) & 15.59/15  (0.41)\\
N/LFG + $F_A^{\rm Deu}$ + hA2018 & 44.21/49  (0.67) & 9.59/13  (0.73) & 10.19/7  (0.18) & 20.89/26  (0.75) & 19.95/15  (0.17)\\
N/LFG + $F_A^{\rm LQCD}$ + INCL & 56.01/49  (0.23) & 9.09/13  (0.77) & 5.75/7  (0.57) & 17.66/26  (0.89) & 26.94/15  (0.03)\\
SF + $F_A^{\rm Deu}$ + hA2018 & 43.52/49  (0.69) & 11.91/13  (0.54) & 11.12/7  (0.13) & 30.98/26  (0.23) & 24.93/15  (0.05)\\
SF + $F_A^{\rm LQCD}$ + INCL\footnote{Best theory tune.} & 66.61/49  (0.05) & 15.72/13  (0.26) & 5.47/7  (0.60) & 23.91/26  (0.58) & 32.25/15  (0.01)\\
N/LFG + $F_A^{\rm Deu}$ + INCL & 68.42/49  (0.03) & 16.33/13  (0.23) & 8.73/7  (0.27) & 27.98/26  (0.36) & 31.56/15  (0.01)\\
SF + $F_A^{\rm Deu}$ + INCL & 81.91/49  (0.00) & 23.94/13  (0.03) & 10.01/7  (0.19) & 35.08/26  (0.11) & 38.83/15  (0.00)\\
\hline
MicroBooNE Tune & 42.61/49 (0.73)  & 7.31/13 (0.89)  & 3.96/7 (0.78)  & 24.59/26 (0.54)  & 17.38/15 (0.30) \\
AR23 & 38.05/49 (0.87)  & 10.63/13 (0.64)  & 7.56/7 (0.37)  & 28.96/26 (0.31)  & 24.07/15 (0.06) \\
    \end{tabular}
\end{ruledtabular}
\end{table*}

Table~\ref{tab:chisq} reports goodness-of-fit metrics ($\chi^2$ and $p$-value)
for all comparisons in this work. Our custom GENIE models appear in
order of increasing sum of $p_\mu$ and $p_p$ $\chi^2$ values. (The impact of
correlations between these two distributions is
small~\cite{MicroBooNE:2024yzp}.) The more empirical FSI and the LQCD axial form factor
give the best results in most cases.
In addition, the Nieves \textit{et al.} and SF QE models often have similar $\chi^2$ values for a given choice of FSI model and axial form factor.
When paired with the lattice QCD axial form factor and the hA2018 FSI
model, these are the best fit combinations. Figure~\ref{fig:theory} includes the pairing with the Nieves/LFG QE model (blue).

\begin{figure*}[htbp]

\centering
\begin{minipage}{0.3\textwidth}
\begin{overpic}[width=\linewidth]{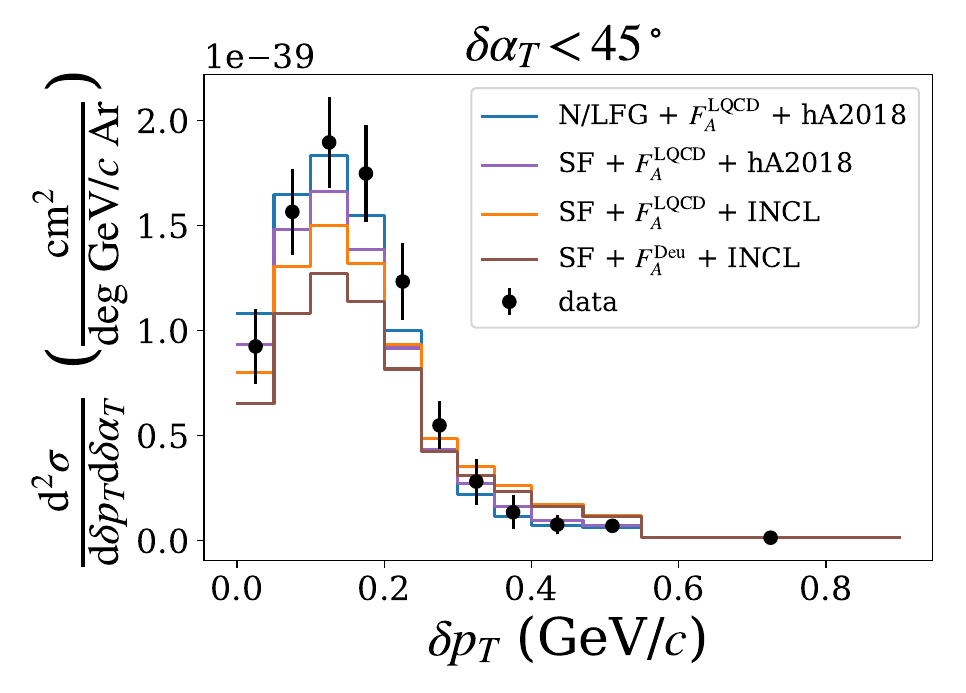}
\put(22,58){\textbf{(a)}}
\end{overpic}
\end{minipage}
\begin{minipage}{0.3\textwidth}
\begin{overpic}[width=\linewidth]{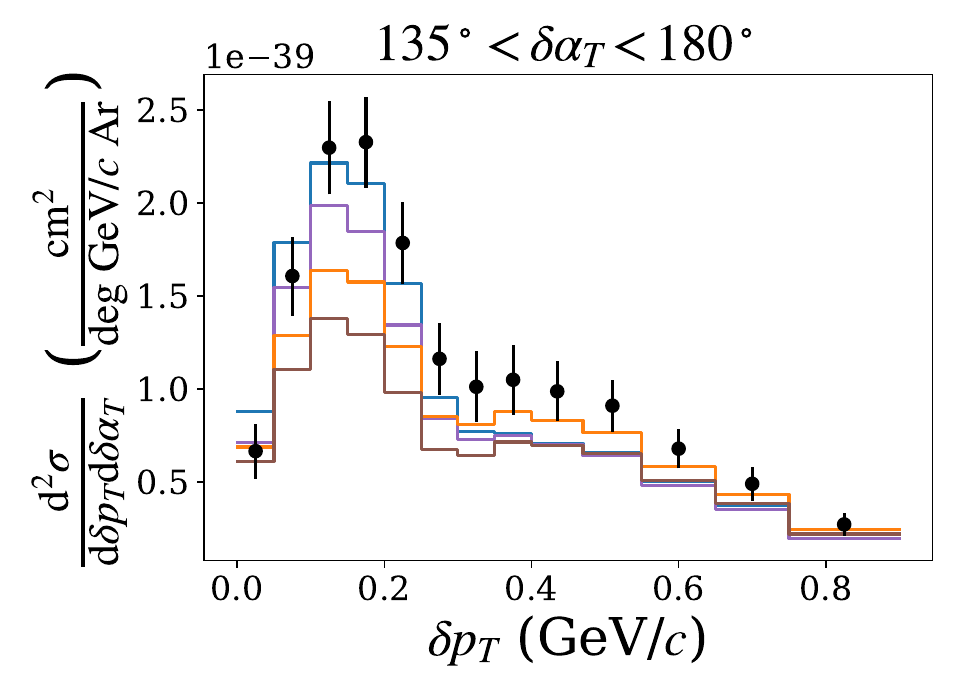}
\put(22,58){\textbf{(b)}}
\end{overpic}
\end{minipage}
\begin{minipage}{0.3\textwidth}
\begin{overpic}[width=\linewidth]{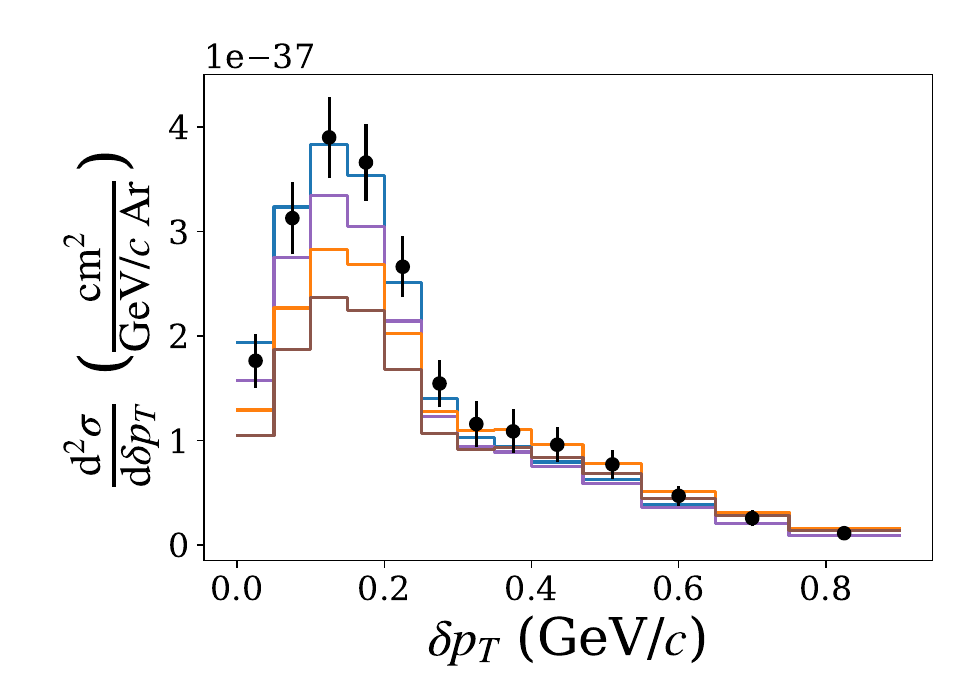}
\put(22,58){\textbf{(c)}}
\end{overpic}
\end{minipage}

\begin{minipage}{0.3\textwidth}
\begin{overpic}[width=\linewidth]{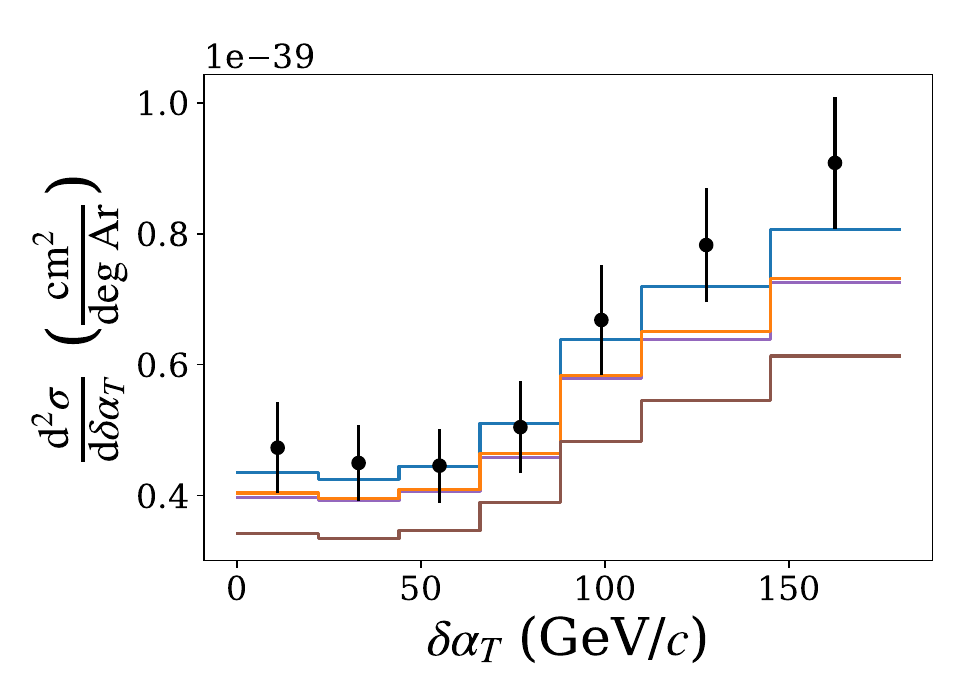}
\put(22,58){\textbf{(d)}}
\end{overpic}
\end{minipage}
\begin{minipage}{0.3\textwidth}
\begin{overpic}[width=\linewidth]{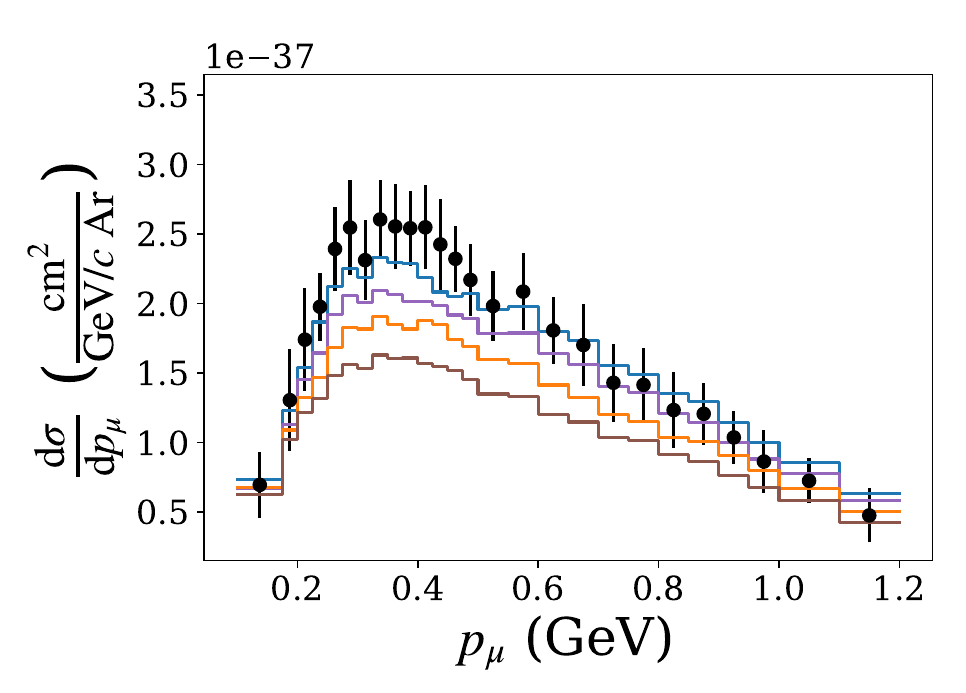}
\put(22,58){\textbf{(e)}}
\end{overpic}
\end{minipage}
\begin{minipage}{0.3\textwidth}
\begin{overpic}[width=\linewidth]{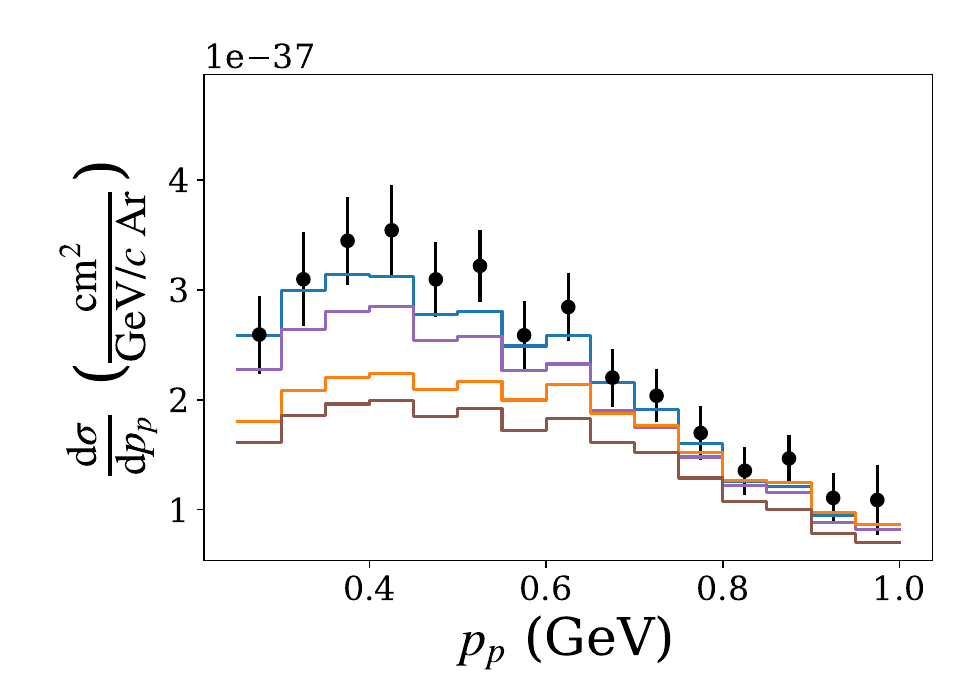}
\put(22,58){\textbf{(f)}}
\end{overpic}
\end{minipage}

\caption{Comparisons with MicroBooNE data~\cite{MicroBooNE:2023tzj, MicroBooNE:2023cmw, MicroBooNE:2024yzp}. Primary comparison is with the empirical model (blue) and various alternate models are included.  See text and Fig.~\ref{fig:theory} for details.}
\label{fig:empirical}
\end{figure*}

In Fig.~\ref{fig:empirical}, the progression starts from the model set
with the best goodness-of-fit (blue, N/LFG+$F_A^{\rm LQCD}$+hA2018). Replacing the QE/nuclear model with SF (purple) yields similar $\chi^2$ values. From there, switches are made
serially toward models that tend to make the agreement with data worse.  First,
the INCL FSI model (orange) is used, and finally $F_A^{\rm Deu}$ (brown).  This
progression leads to a clear separation of models.  Predictions using INCL FSI and $F_A^{\rm Deu}$
give decidedly worse matches to the data.

For comparison between the new GENIE configurations and those used by current
experiments, the last two rows of Table~\ref{tab:chisq} include $\chi^2$ scores
and $p$-values for the MicroBooNE Tune and AR23. (Plots are in the Supplemental Material~\cite{suppl}.)
The most theoretical model and AR23 show similar discrepancies with the data.
The MicroBooNE tune prediction is similar to the best fit result.

Thanks to GENIE's flexible configuration, each model component can be tested
with any combination of the other choices.  The main observation from our
comparisons is that the models containing 
hA2018 and $F_A^{\rm LQCD}$ with either N/LFG or SF result in the best fit and match the MicroBooNE data well ($\chi^2$/bin less than 1).  On the other hand, the most theoretical model (SF+INCL+$F_A^{\rm LQCD}$)
produces an unsatisfactory fit to that data.

The $\chi^2$ values of Nieves/LFG and SF QE/nuclear models are typically similar for all combinations of FSI and axial form factor models, suggesting that
the data have limited sensitivity to details of the nuclear ground state. For
FSI, the hA2018 model has $\chi^2$ values consistently better than INCL.
Although $\chi^2$ weights the tail of $p_T$ more than the peak, the peak is
clearly influenced by many factors and the best fit manages to match both peak
and tail.

Notably, $F_A^{\rm LQCD}$ consistently gives a better fit to MicroBooNE data
than $F_A^{\rm Deu}$, which is based on neutrino-deuterium
data~\cite{Meyer:2016oeg} that have been the basis of simulations for many years.  This conclusion is of course more complicated because MEC models
also have a similar affect on normalization as the axial form factors.  GENIE has implemented SuSAv2~\cite{Megias:2018ujz},
Valencia~\cite{Gran:2013kda}, and Martini~\cite{Martini:2010ex} \textit{et al.}
MEC models. The SuSAv2 calculation~\cite{Megias:2018ujz} is
used in this work for all predictions except the MicroBooNE Tune.

Using an average over the two alternative MEC treatments produces differences in the GENIE predictions for the MicroBooNE distributions under study that are significantly smaller (approximately $1/3$ as large) as those caused by the change from $F_A^{\rm Deu}$ to $F_A^{\rm LQCD}$~\cite{suppl}.
For simplicity and clarity in the results, we therefore omitted MEC variations from our comparisons. 

In conclusion, GENIE's especially flexible code structure allows
straightforward interchange of model components in comparisons with data. Recent
improvements to GENIE involve both a correction to the implementation of
the Nieves \textit{et al.} QE cross section~\cite{suppl} as well as several new models. We show comparisons with MicroBooNE neutrino-argon CC$0\pi$ data including protons~\cite{MicroBooNE:2023cmw, MicroBooNE:2023tzj,
MicroBooNE:2024yzp} using various combinations of empirical or often-used model configurations -- Nieves/LFG, hA2018, or $F_A^{\rm
Deu}$ -- with more theoretical
alternatives -- SF, INCL, and $F_A^{\rm LQCD}$.
These tests provide a detailed benchmark for upcoming
experiments~\cite{MicroBooNE:2015bmn, DUNE:2020lwj} which also use an argon
target.  This study is similar to that of Filali, Munteanu, and
Dolan~\cite{Filali:2024vpy} and Nikolakopoulos \textit{et al.}~\cite{Nikolakopoulos:2024mjj}, but extensions to a more inclusive data set~\cite{MicroBooNE:2024yzp}, more complete models, and newer models are made. The best-performing models studied here consequently 
achieve significantly better agreement with MicroBooNE data.

Goodness-of-fit quality in Table \ref{tab:chisq} ranges from excellent ($\chi^2$
per degree of freedom less than 1) to poor ($p$-value less than 0.05). 
These data are not able to distinguish between the LFG or SF nuclear models.  Although other neutrino studies see moderate improvements with the SF nuclear  model~\cite{Filali:2024vpy,Nikolakopoulos:2024mjj}, an examination of $(e,e'p)$ data~\cite{CLAS:2025fqh} shows a clear preference for the SF model. On the other hand, our MicroBooNE comparisons reveal a
preference for the LQCD axial form factor over the $\nu$-$\isotope[2]{H}$ based
fit and the hA2018 FSI model over INCL. In particular, the $p_p$ and
$\delta p_T\ vs.\ \delta \alpha_T$ distributions yield $p$-values less than 0.05 for model sets that include both $F_A^{\rm
Deu}$ and INCL.

The best variations of the core GENIE models presented here provide a better
fit to the MicroBooNE data than either the AR23 or the MicroBooNE Tune reference model
sets.  It is interesting to note that a primary result from the empirical
MicroBooNE~\cite{MicroBooNE:2015bmn} and GENIE~\cite{GENIE:2022qrc} tunes to hydrocarbon data is the need to enhance the QE cross section magnitude beyond
the commonly used models~\cite{NievesQE, SuSAv2QE}. 
The data preference for $F_A^{\rm LQCD}$ seen in
this work provides a reasonable way to produce that QE enhancement even for a heavier target (argon).


This work indicates that the choice of axial form factor is the
most important contribution to $\chi^2$ improvement.  This is the first quantitative preference for the LQCD form factor demonstrated using MicroBooNE data.  Although LQCD calculations of the axial QE form factor are a
poor fit to neutrino-deuterium data, both $\nu ^2H$ data fits and LQCD calculations provide a reasonable fit to MINERvA axial form factor data~\cite{MINERvA:2025ygc}.
Other components will contribute to QE-like simulations and can affect this assessment.  One of the biggest is the choice of MEC model which we show is a comparatively small effect~\cite{suppl}.
The choice of QE/nuclear model is a lesser effect.  The SF model is based on fits to electron scattering data and is
among the best fits to the MicroBooNE neutrino data here.

For FSI, the hA2018 model is consistently better than INCL.  Ref.~\cite{Nikolakopoulos:2024mjj} also studied the role of FSI in the MicroBooNE TKI data and finds similar results for the INCL model.  This is
surprising for a few reasons.  The hA2018 model is primarily based on
hadron-nucleus data and simple models of the final state kinematics.  However,
the proton total channel-dependent cross section (e.g., total spallation) values
depend heavily on a model~\cite{Mashnik:2005ay}, and hA2018 lacks Pauli blocking and other features present in modern FSI models. The INCL model includes a variety of effects beyond what is available in hA2018, e.g., a nuclear model that has been developed to fit proton-nucleus scattering data.
A key feature of INCL is the depletion of protons emitted due to nucleon cluster formation; this subject has been previously discussed~\cite{Ershova:2022jah}, but cluster emission data in neutrino experiments are required to understand implications for data analysis and simulation.

Although TKI variables have been specifically designed to probe nuclear
effects, we nevertheless find that more traditional observables have similar
sensitivity for the model variations and MicroBooNE data sets studied. In this
work, the $\delta p_T$ and $p_\mu$ distributions both provide QE/nuclear model
discrimination, while both $\delta \alpha_T$ and $p_p$ do the same for FSI.

Although these comparisons to MicroBooNE data provide important insight into GENIE's
present performance, future work can be more comprehensive. Additional model
constraints could come from comparisons to neutrino data with
different beam energies and nuclear targets. Theoretical investigation of
relevant nuclear effects (e.g., RPA correlations) is still ongoing. 
Valuable comparisons
can also be made with electron and hadron scattering data.  We
look forward to improvements in modeling and fits to data inside GENIE and elsewhere.

The authors acknowledge helpful discussions with Jean-Christophe David (Saclay) on the INCL code. We also thank Stephen Dolan for feedback on an early draft of this work. This manuscript has been authored by Fermi Forward Discovery Group, LLC under
Contract No. 89243024CSC000002 with the U.S. Department of Energy, Office of
Science, Office of High Energy Physics.

\bibliographystyle{apsrev4-2}
\bibliography{references}

\end{document}


\title{Supplemental Materials for Benchmarking State-of-the-Art Theory and Empirical Models of Pionless Neutrino-Argon Scattering in GENIE}

\maketitle

\section{Correction to GENIE implementation of the Nieves QE model}

The Nieves \textit{et al}. quasielastic (QE) cross-section model~\cite{Nieves:2004wx} was first implemented in GENIE v2.12.0 as a non-default simulation component. With the introduction of multiple official configurations in the GENIE v3 release series, it was included in several widely-used model sets, including G18\_10a\_02\_11a (the basis for the MicroBooNE Tune~\cite{MicroBooNE:2021ccs}) and AR23\_20i\_00\_000 (used by SBN and DUNE).

In the original calculation from Ref.~\cite{Nieves:2004wx}, the differential cross section is proportional to a contraction of leptonic ($L_{\mu\nu}$) and nuclear ($W^{\mu\nu}$) tensors, which are defined in terms of matrix elements of the relevant weak currents. This tensor contraction is greatly simplified by working in the laboratory frame (where the target nucleus is at rest) and assigning coordinates so that the 3-momentum transfer $\vec{q}$ points along the positive $z$ direction. Under these choices, the full tensor contraction may be written in terms of only five elements of the nuclear tensor: $W^{00}$, $W^{xx}$, $W^{zz}$, $W^{xy}$, and $W^{0z}$ (where the index $0$ denotes the time component). Since only QE scattering is of interest, $W^{\mu\nu}$ is computed by integrating a single-nucleon tensor $A^{\mu\nu}$ (which contains the nucleon matrix elements) over the nuclear volume and the local Fermi gas (LFG) momentum distribution. The elements of these two tensors correspond, e.g., $W^{xx}$ is computed by integrating $A^{xx}$. Random Phase Approximation (RPA) corrections for long-range nucleon-nucleon correlations are handled in the model by modifying $A^{\mu\nu}$ from the standard expression used for a free nucleon. GENIE allows these corrections to be toggled on and off.

While the original Nieves \textit{et al.} model made predictions for the outgoing lepton only, an event generator must describe all details of each interaction. In GENIE, the simulation is done in the nucleon rest frame, and the integrals used to compute $W^{\mu\nu}$ are replaced with Monte Carlo sampling of the initial nucleon position and momentum. For a given nucleon sampled in this way, the differential cross section then becomes proportional to the contraction $L_{\mu\nu}A^{\mu\nu}$, and the outgoing nucleon kinematics may be inferred from the final lepton via conservation of 4-momentum.

When adding the Nieves \textit{et al.} model, GENIE authors implemented the simplified form of the tensor contraction mentioned above, which requires calculation of only five elements of $A^{\mu\nu}$. However, because the struck nucleon is not initially at rest in the lab frame (unlike the target nucleus), the assumptions required for the simplification are violated. As a result, the implementation of this model in all public releases of GENIE (including the current v3.6.2) underestimates the QE cross section.

To address this problem, we have adjusted the GENIE code so that the evaluation of $L_{\mu\nu}A^{\mu\nu}$ occurs in the rest frame of the initial nucleon. Coordinates are rotated within this frame to ensure that $\vec{q}$ points along $+z$. Under these conditions, the simplified form of the tensor contraction may be used. When RPA corrections are disabled in GENIE (as is done for our model comparisons in the main text), this change of frame is sufficient to fully resolve the implementation error. However, the non-relativistic treatment of RPA in Ref.~\cite{Nieves:2004wx} (and therefore in GENIE) is not fully Lorentz invariant. As a result, a fix compatible with the RPA corrections has not yet been completed and users should run the Nieves \textit{et al.} QE model without it.

\begin{figure}[htbp]

\centering

\begin{minipage}{0.45\textwidth}
\begin{overpic}[width=\linewidth]{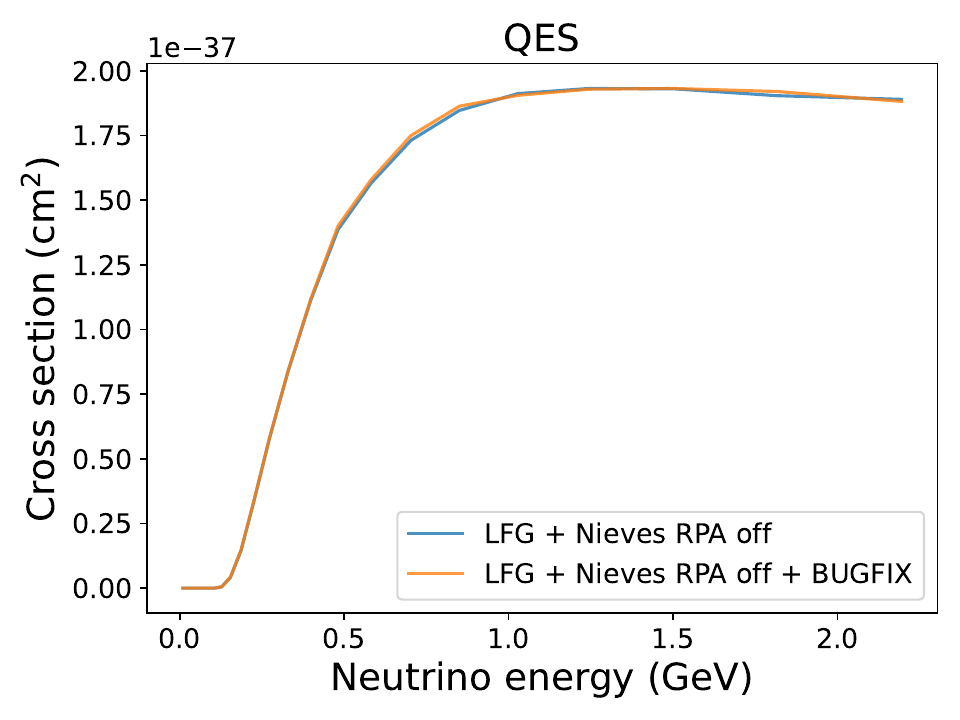}
\put(22,58){\textbf{(a)}}
\end{overpic}
\end{minipage}
\begin{minipage}{0.45\textwidth}
\begin{overpic}[width=\linewidth]{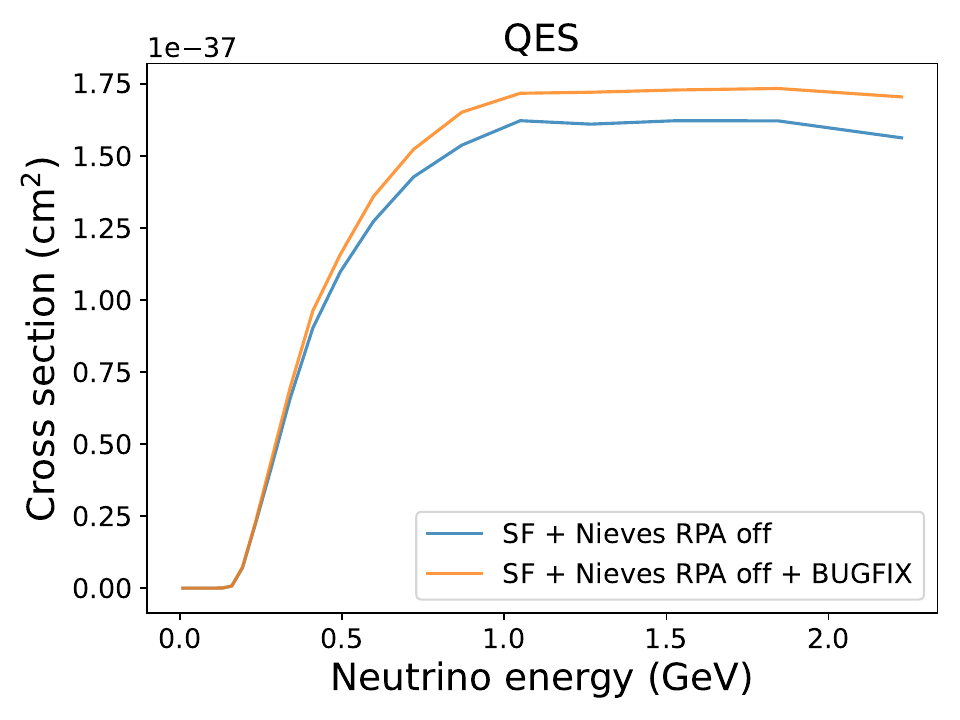}
\put(22,58){\textbf{(b)}}
\end{overpic}
\end{minipage}



\caption{Impact of the error in the implementation of the Nieves \textit{et al.} QE model in GENIE.}
 \label{fig:spline}
\end{figure}

In Fig.~\ref{fig:spline}, we assess the impact of the implementation error on GENIE's prediction for the total QE cross section $\sigma(E_\nu)$ as a function of neutrino energy $E_\nu$. All results shown here are obtained with RPA corrections turned off. In Fig.~\ref{fig:spline}(a), the cross sections obtained using the LFG nuclear model are almost the same whether our fix is applied (orange) or not (blue). In Fig.~\ref{fig:spline}(b), which shows similar results for the spectral function (SF) nuclear model, the fix corrects a deficit of about 8\%. The implementation problem becomes more noticeable for the SF due to its inclusion of very high-momentum nucleons above the Fermi momentum. The rest frames of these nucleons differ from the lab frame most significantly.

\section{Effect of RPA corrections on GENIE predictions for MicroBooNE}

\begin{figure}[htbp]

\centering

\begin{minipage}{0.45\textwidth}
\begin{overpic}[width=\linewidth]{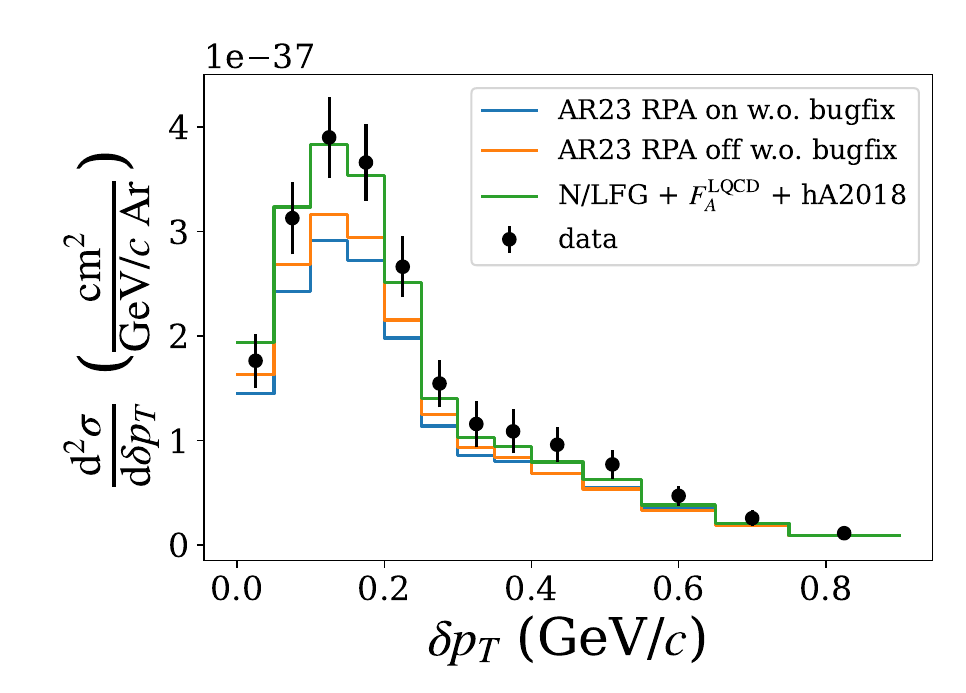}
\put(22,58){\textbf{(a)}}
\end{overpic}
\end{minipage}
\begin{minipage}{0.45\textwidth}
\begin{overpic}[width=\linewidth]{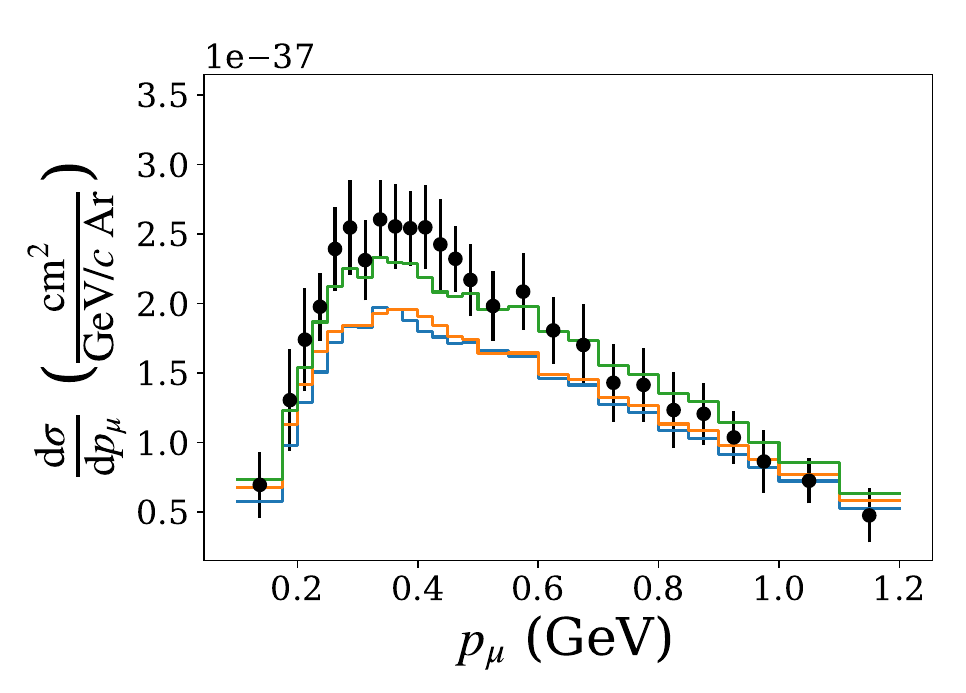}
\put(22,58){\textbf{(b)}}
\end{overpic}
\end{minipage}

\caption{Impact of the Nieves \textit{et al}.\ RPA corrections on GENIE predictions of the MicroBooNE $\delta p_T$ and $p_{\mu}$ measurements.}
 \label{fig:rpa_effect}
\end{figure}

Since the recipe for RPA corrections in the Nieves \textit{et al.} model is based on an LFG treatment of the nucleus, it is incompatible with the alternative SF nuclear model discussed in the main text. To perform comparisons to the MicroBooNE data with both nuclear models while maintaining theoretical consistency, we therefore omit RPA corrections in all of our custom GENIE configurations. These corrections are known to significantly reduce the cross section at low $Q^2$ or, equivalently, the most forward scattering angles. For the MicroBooNE measurements considered in this work, however, the impact of RPA is modest and sub-leading compared to some other effects. In Fig.~\ref{fig:rpa_effect}, we compare GENIE predictions for MicroBooNE using the AR23\_20i\_00\_000 model set (orange, RPA corrections enabled) to otherwise equivalent predictions with RPA turned off (blue). Only two MicroBooNE measurements are shown, but the trends seen here are representative of all data sets under study. Switching to the lattice QCD axial form factor (green, RPA off) is seen to have a much stronger influence on the level of agreement with the data.

\section{Impact of the MEC model choice on our comparisons to MicroBooNE data}

Multiple processes contribute to the QE-like signal of the MicroBooNE measurements. Although we omit MEC model variations from the main text, they were also examined in our study. Currently, there are three theoretical models of MEC interactions available in GENIE -- SuSAv2~\cite{Dolan:2019bxf}, Valencia~\cite{Gran:2013kda}, and Martini \textit{et al.}~\cite{Russo:2025oph}. All of these are implemented via a factorization approach. Tables of precomputed nuclear response functions are used to calculate inclusive cross sections from which lepton kinematics are sampled. The final-state nucleons are then simulated using a simple nucleon cluster model~\cite{Katori:2013eoa}.

This simulation workflow facilitates inclusion of new models and is commonly used in neutrino event generators. However, predictions for hadronic observables can noticeably differ from a fully unfactorized calculation~\cite{Nikolakopoulos:2023pdw}. Especially because of the precomputed nuclear responses, physics inconsistencies can also easily arise between these MEC implementations and other GENIE components. In particular, the definitions of the QE (1p1h) and MEC (2p2h) channels must be consistent with each other for a truly valid simulation, but details can differ between theoretical treatments.  For example, contributions from short-range nucleon-nucleon correlations are assigned to QE within the SuSAv2 formalism and to MEC by Martini \textit{et al.}~\cite{Russo:2025oph}.
In addition, the changes we make in axial form factor between $F_A^\mathrm{Deu}$ and $F_A^\mathrm{LQCD}$ should have an effect on both QE and MEC simulations, but current GENIE versions only account for the impact on QE.




Inconsistencies like these are common in neutrino event generators due to the formidable requirement to comprehensively model all relevant interaction physics for experiments. Achieving fully consistent calculations that meet experimental needs is very difficult and beyond the scope of the present work. Rather than attempt to resolve all potential incompatibilities between the available GENIE MEC treatments and our other model components, we instead estimate the overall impact of MEC model variations on our MicroBooNE comparisons using a quantitative metric.

Our starting GENIE configuration is the best-fit empirical model (N/LFG+$F_A^{\rm LQCD}$+hA2018) from the main text, which we abbreviate as B.E. This model set uses SuSAv2 to simulate MEC interactions. We then produce alternative configurations that are identical except that SuSAv2 MEC is replaced with either the Valencia or the Martini \textit{et al.} treatment.

\begin{figure}[htbp]
\centering
\begin{minipage}{0.3\textwidth}
\begin{overpic}[width=\linewidth]{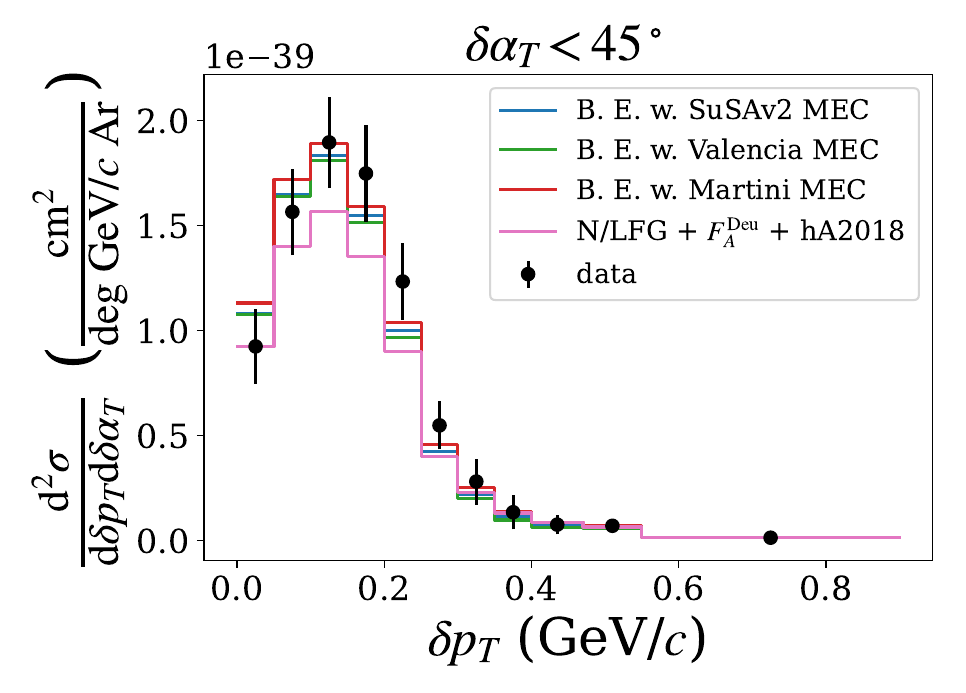}
\put(22,58){\textbf{(a)}}
\end{overpic}
\end{minipage}
\begin{minipage}{0.3\textwidth}
\begin{overpic}[width=\linewidth]{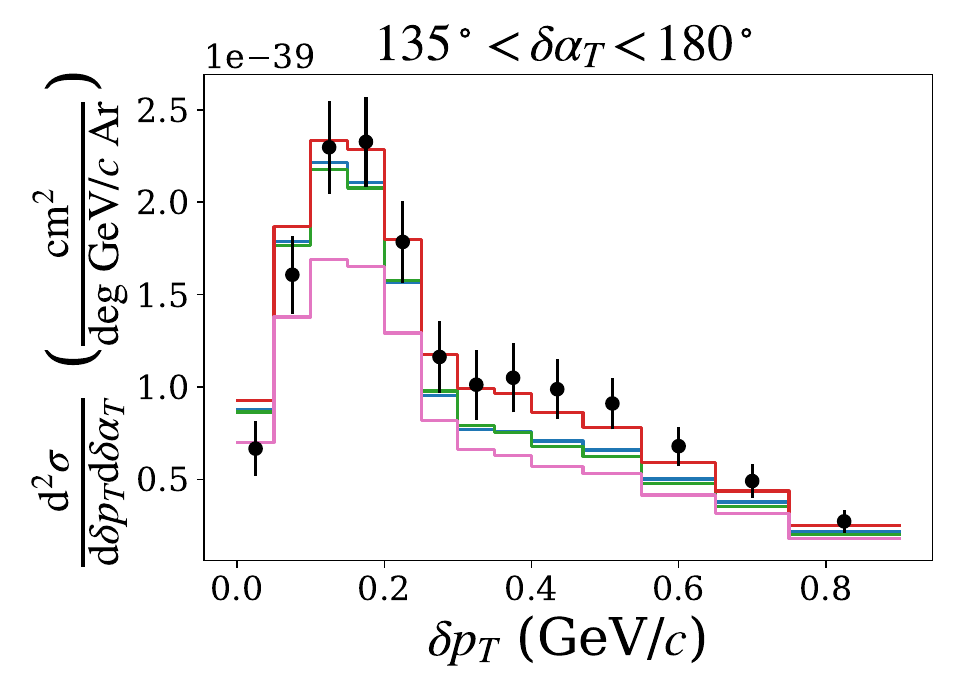}
\put(22,58){\textbf{(b)}}
\end{overpic}
\end{minipage}
\begin{minipage}{0.3\textwidth}
\begin{overpic}[width=\linewidth]{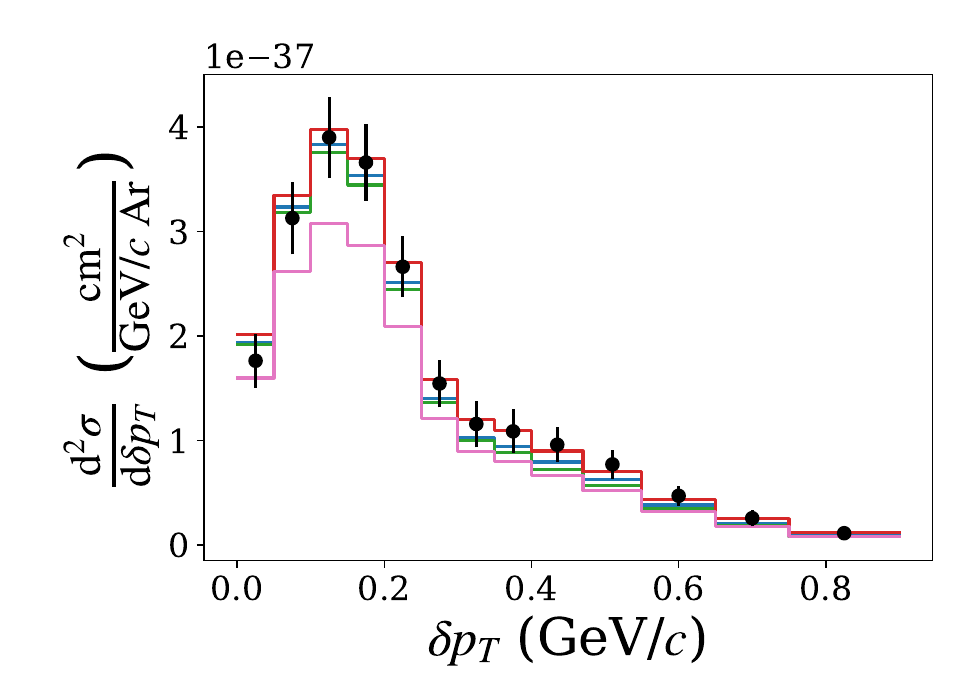}
\put(22,58){\textbf{(c)}}
\end{overpic}
\end{minipage}

\begin{minipage}{0.3\textwidth}
\begin{overpic}[width=\linewidth]{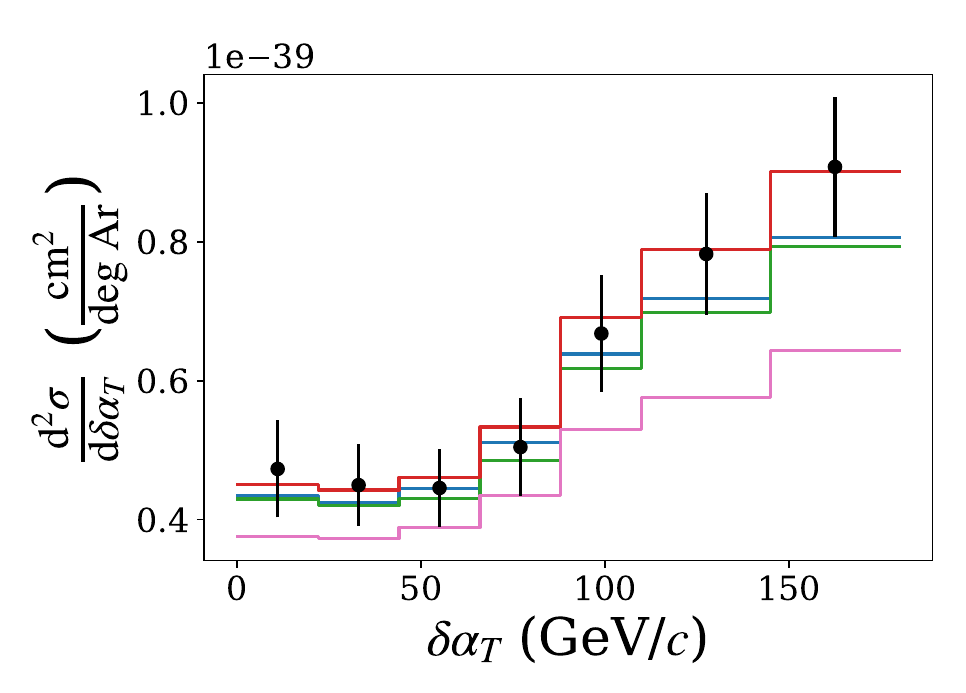}
\put(22,58){\textbf{(d)}}
\end{overpic}
\end{minipage}
\begin{minipage}{0.3\textwidth}
\begin{overpic}[width=\linewidth]{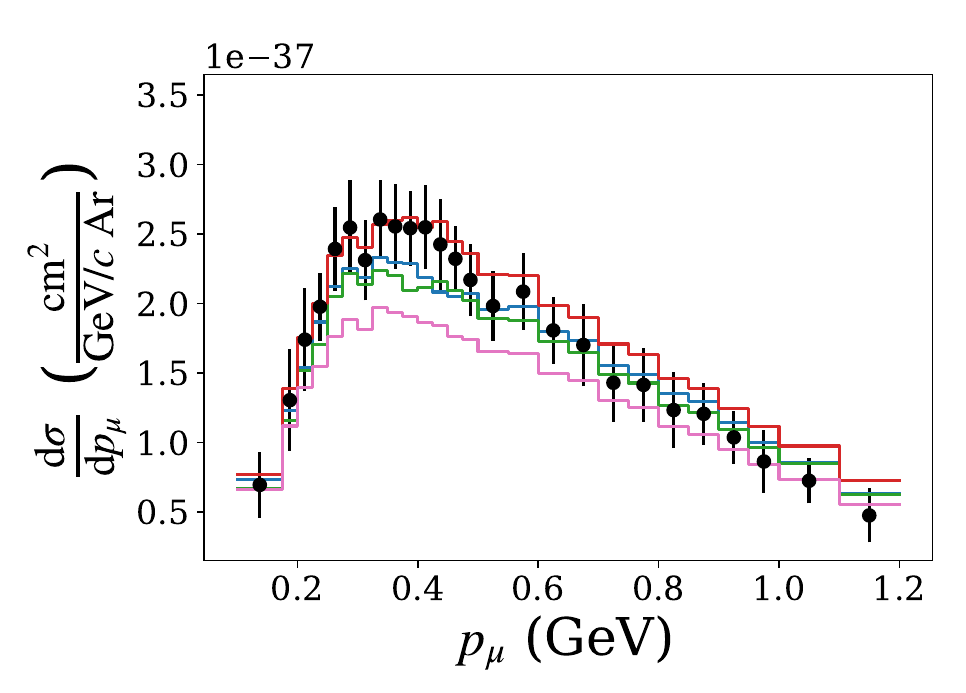}
\put(22,58){\textbf{(e)}}
\end{overpic}
\end{minipage}
\begin{minipage}{0.3\textwidth}
\begin{overpic}[width=\linewidth]{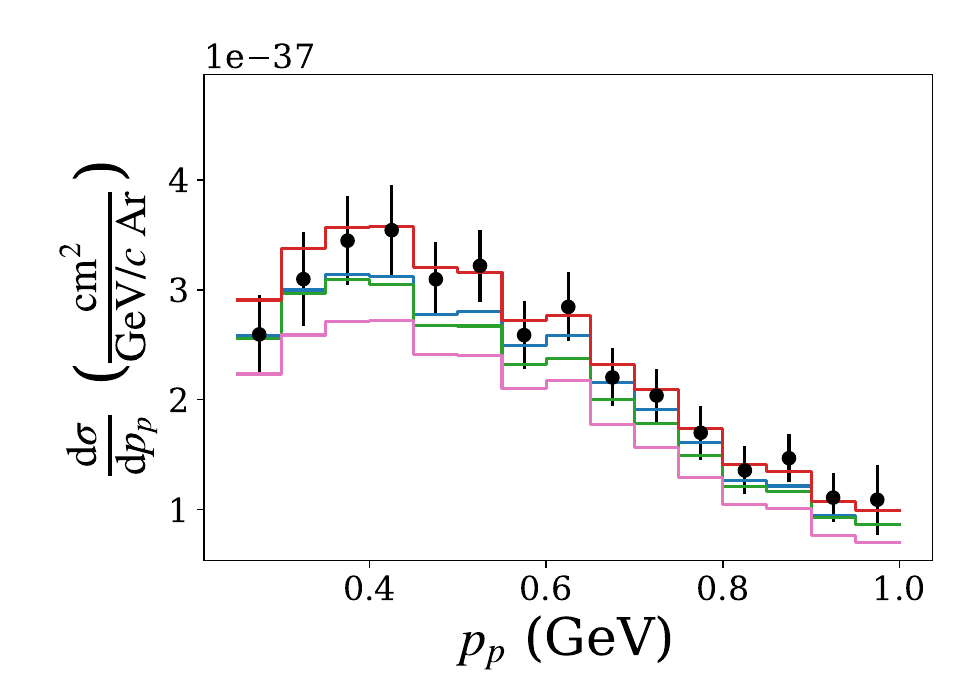}
\put(22,58){\textbf{(f)}}
\end{overpic}
\end{minipage}

\caption{Comparisons with MicroBooNE data~\cite{MicroBooNE:2023tzj, MicroBooNE:2023cmw, MicroBooNE:2024yzp}. The reference B.E. configuration uses the SuSAv2 MEC model (blue). Variations are shown in which the Valencia (green) or Martini \textit{et al.} (red) MEC models are used instead. The pink line shows a different variation in which SuSAv2 MEC is kept and $F_A^{\rm LQCD}$ is replaced with $F_A^{\rm Deu}$.}
 \label{fig:best_empirical_mec}
\end{figure}

\begin{table*}[htbp]

    \caption{$\chi^2/{\rm ndf}$ ($p$-value)  of MC and reference simulations relative to MicroBooNE data. }
    \label{tab:chisq_mec}
    \centering
\begin{ruledtabular}
    \begin{tabular}{lrrrrr}
    Configuration & $\delta p_T$ vs. $\delta \alpha_T$ & $\delta p_T$ & $\delta \alpha_T$ & $p_{\mu}$ & $p_p$ \\
    \hline
B.E. (N/LFG+$F_A^{\rm LQCD}$+hA2018) & 35.98/49  (0.92) & 6.78/13  (0.91) & 3.47/7  (0.84) & 16.29/26  (0.93) & 16.05/15  (0.38)\\
B.E. with Valencia MEC & 37.55/49  (0.88) & 7.89/13  (0.85) & 3.28/7  (0.86) & 15.52/26  (0.95) & 17.52/15  (0.29)\\
B.E. with Martini \textit{et al.} MEC & 32.50/49  (0.97) & 3.99/13  (0.99) & 1.43/7  (0.98) & 10.29/26  (1.00) & 11.42/15  (0.72)\\
N/LFG + $F_A^{\rm Deu}$ + hA2018 & 44.21/49  (0.67) & 9.59/13  (0.73) & 10.19/7  (0.18) & 20.89/26  (0.75) & 19.95/15  (0.17)\\
    \end{tabular}
\end{ruledtabular}
\end{table*}

Figure~\ref{fig:best_empirical_mec} shows comparisons to the MicroBooNE data (same as is shown in main article) for the B.E. configuration (blue) and with the MEC component replaced by the Valencia (green) and Martini \textit{et al.} (red) predictions. A variation of B.E. in which SuSAv2 MEC is retained and the axial form factor is changed to $F_A^{\rm Deu}$ (pink) is also shown. Using the predicted cross sections plotted in the figure, we compute the root-mean-square fractional deviation
\begin{equation}
\mathcal{F} \equiv \sqrt{ \frac{1}{N} \sum_{j=1}^N
  \bigg( \frac{ a_j - b_j }{ b_j } \bigg)^2} 
\end{equation} from the B.E. prediction for the $F_A^{\rm Deu}$ variation and for the average of the two MEC variations. Here $b_j$ is the B.E. prediction for the differential cross section in the $j$-th bin, and the sum runs over all $N = 110$ bins included in the data sets shown in Fig.~\ref{fig:best_empirical_mec}. For the $F_A^{\rm Deu}$ variation, $a_j$ is the corresponding prediction in the $j$-th bin. For the MEC variations, $a_j$ is the mean of the predictions obtained using the Valencia and Martini \textit{et al.} models instead of SuSAv2. Following this procedure, we obtain $\mathcal{F} = 4.78\%$ for the MEC variations, which is slightly less than one third of the value $\mathcal{F} = 15.6\%$ for the switch to $F_A^{\rm Deu}$. We interpret this result as an indication that the MicroBooNE data are significantly less sensitive to MEC modeling details than to the difference between the $F_A^{\rm Deu}$ and $F_A^{\rm LQCD}$ form factors.

Table~\ref{tab:chisq_mec} lists $\chi^2$ scores and $p$-values for the GENIE predictions shown in Fig.~\ref{fig:best_empirical_mec}. While the B.E. variation with Martini \textit{et al.} MEC notably has the lowest $\chi^2$ values for all of the MicroBooNE measurements studied, the consistently good $p$-values ($> 0.05$) show that the data are unable to fully discriminate between the competing MEC models available in GENIE.

\section{Comparison with tunes used by  experiments}

\begin{figure}[htbp]

\centering
\begin{minipage}{0.3\textwidth}
\begin{overpic}[width=\linewidth]{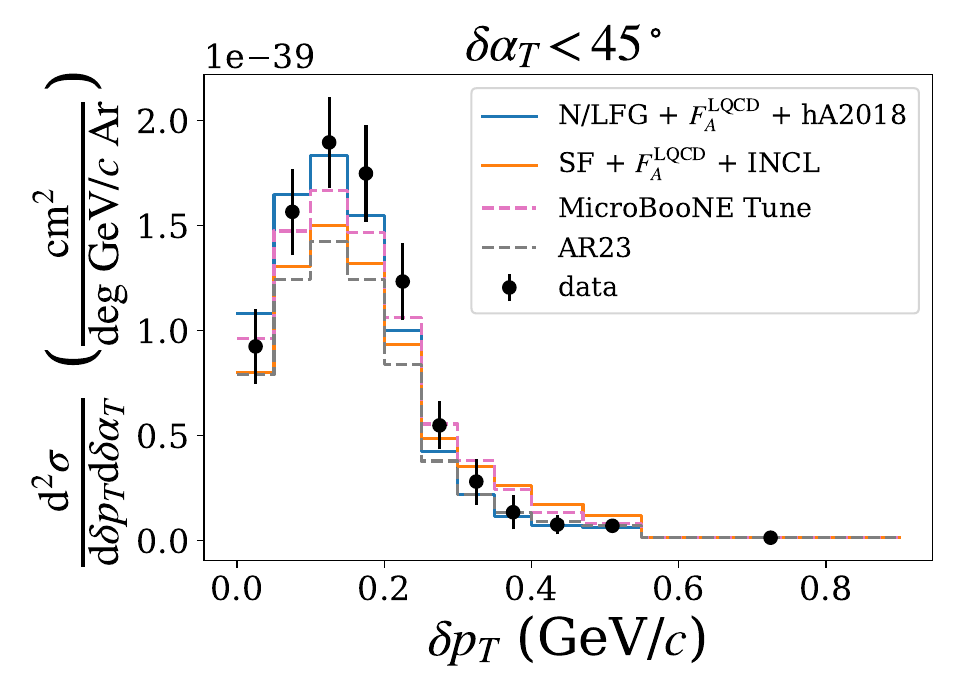}
\put(22,58){\textbf{(a)}}
\end{overpic}
\end{minipage}
\begin{minipage}{0.3\textwidth}
\begin{overpic}[width=\linewidth]{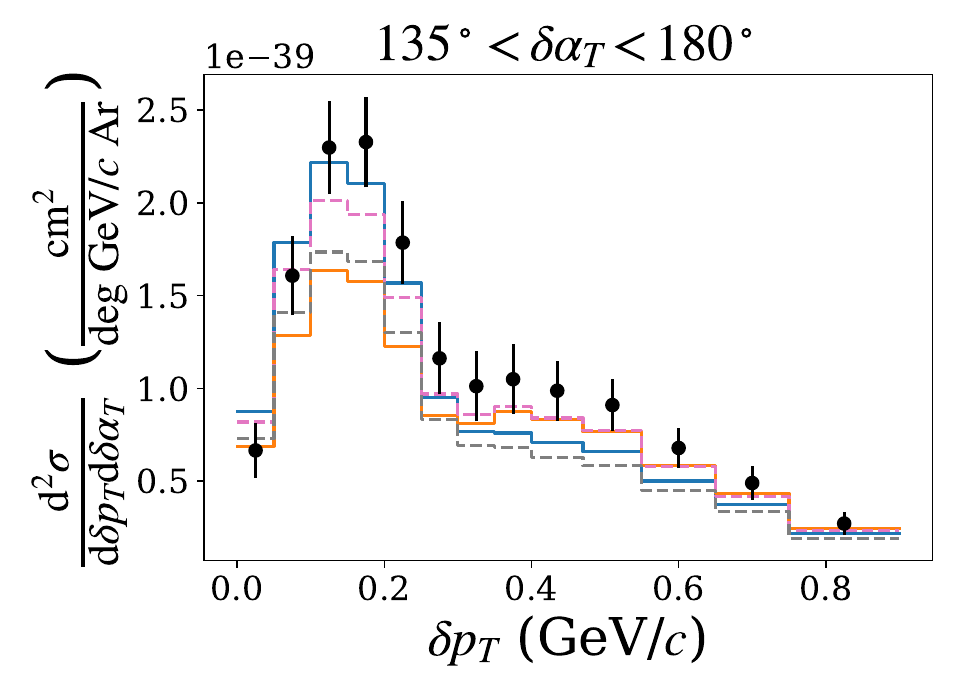}
\put(22,58){\textbf{(b)}}
\end{overpic}
\end{minipage}
\begin{minipage}{0.3\textwidth}
\begin{overpic}[width=\linewidth]{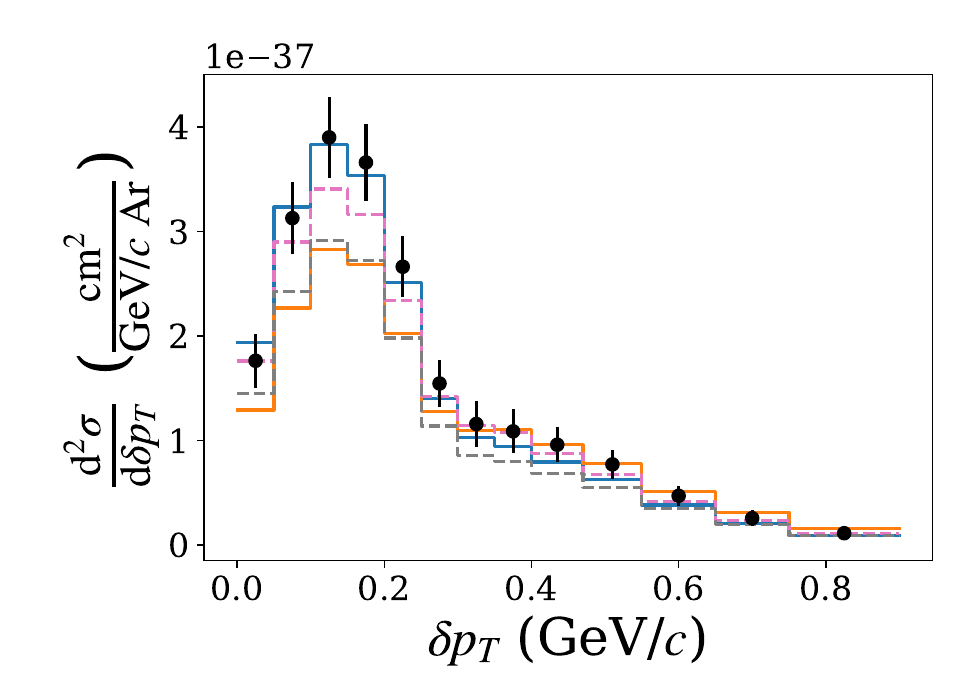}
\put(22,58){\textbf{(c)}}
\end{overpic}
\end{minipage}

\begin{minipage}{0.3\textwidth}
\begin{overpic}[width=\linewidth]{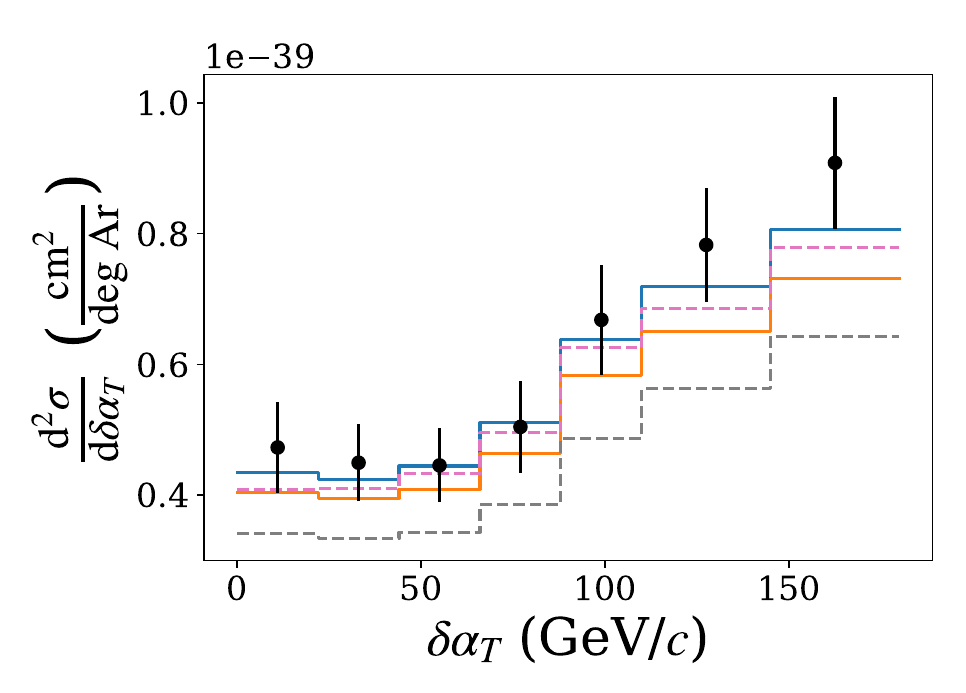}
\put(22,58){\textbf{(d)}}
\end{overpic}
\end{minipage}
\begin{minipage}{0.3\textwidth}
\begin{overpic}[width=\linewidth]{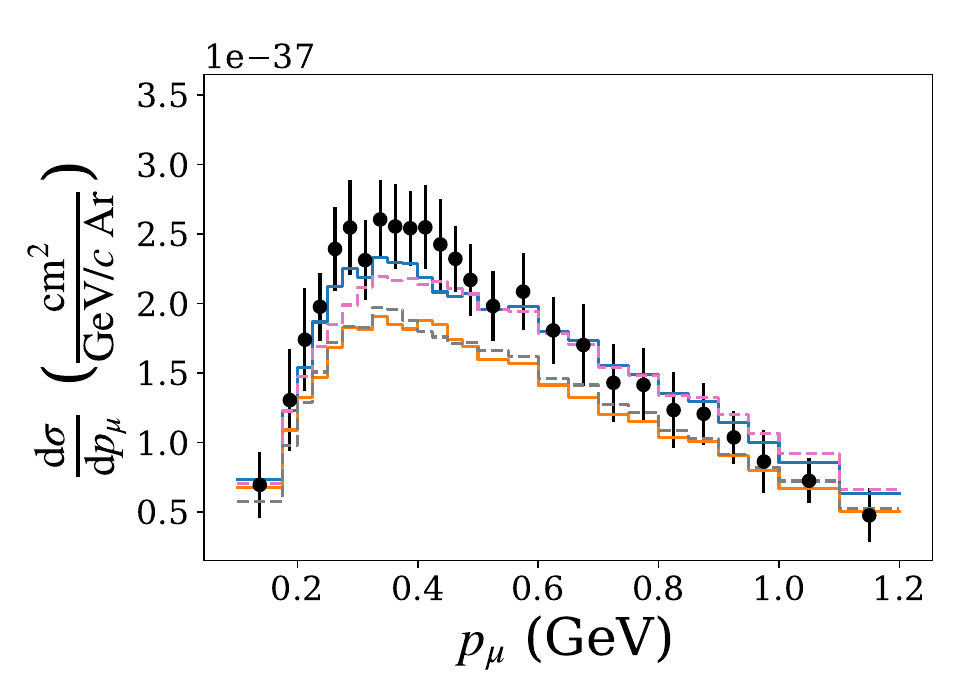}
\put(22,58){\textbf{(e)}}
\end{overpic}
\end{minipage}
\begin{minipage}{0.3\textwidth}
\begin{overpic}[width=\linewidth]{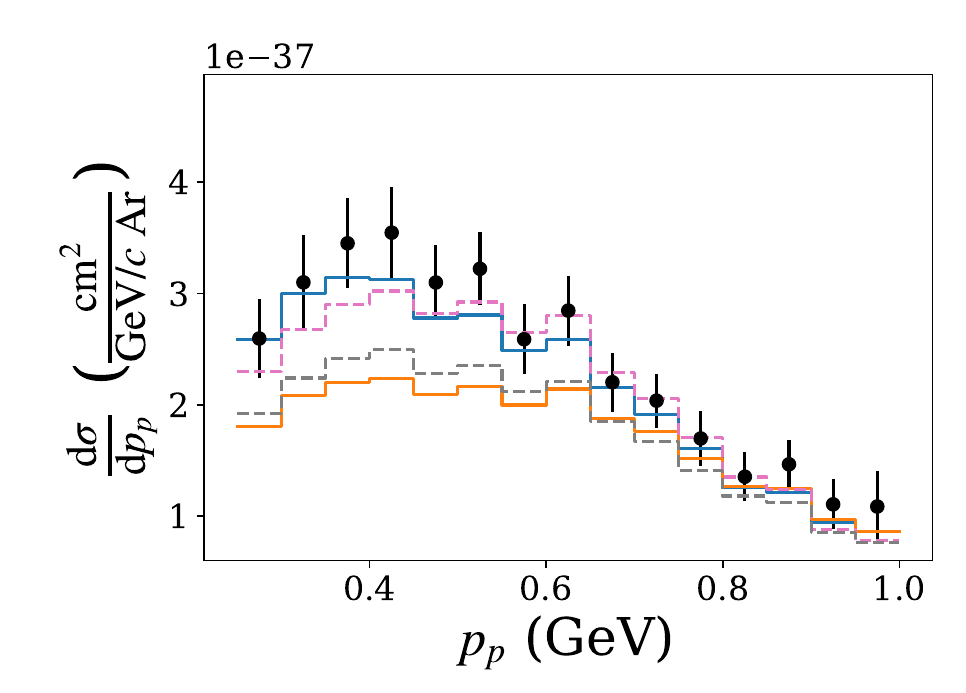}
\put(22,58){\textbf{(f)}}
\end{overpic}
\end{minipage}


\caption{Comparisons of experimental GENIE configurations with the MicroBooNE data~\cite{MicroBooNE:2023tzj, MicroBooNE:2023cmw, MicroBooNE:2024yzp} (black points). The MicroBooNE Tune (dashed pink) and AR23 (dashed gray) predictions are plotted together with those obtained using our B.E. (blue) and M.T. (orange) configurations.}
 \label{fig:uboonetune_ar23}
\end{figure}

To examine how the new GENIE configurations relate to those used by current
experiments, the last two rows of Table~I in the main text include $\chi^2$ scores
and $p$-values for the MicroBooNE Tune and AR23. In Fig.~\ref{fig:uboonetune_ar23}, we provide the corresponding plots that compare these models with the MicroBooNE data discussed in the main text~\cite{MicroBooNE:2023tzj, MicroBooNE:2023cmw, MicroBooNE:2024yzp}. The predictions of the MicroBooNE Tune~\cite{MicroBooNE:2021ccs} and AR23 are plotted together with those of the 
B.E. configuration mentioned above, and we also include our most theoretical (SF+$F_A^{LQCD}$+INCL, abbreviated M.T.) GENIE prediction. The B.E. model and the MicroBooNE Tune are both good fits to the data, but the latter has a slightly worse normalization. The M.T. model and AR23 have similar discrepancies with the data.

\bibliographystyle{apsrev4-2}
\bibliography{references}